\documentclass[aps,prb,twocolumn,showpacs,amsmath,amssymb]{revtex4}
\usepackage{epsfig}
\usepackage{bm}

\begin{document}
\title{Statistics of Charge Fluctuations in Chaotic Cavities}
\author{S. Pilgram and M. B\"uttiker}
\affiliation{D\'epartement de Physique Th\'eorique, Universit\'e de Gen\`eve,
        CH-1211 Gen\`eve 4, Switzerland}
\date{\today}

\begin{abstract}
We consider the zero frequency
fluctuations of charge inside a mesoscopic conductor in
the large capacitance limit. In analogy
to current counting statistics we derive the characteristic
function of charge fluctuations in terms of the scattering matrix
of the conductor.
Using random matrix theory
we evaluate the characteristic function 
semi-analytically for chaotic cavities.
Our result is universal in the sense that it describes
not only the fluctuations of charge, but of
any observable quantity inside the
cavity. We discuss equilibrium and
non-equilibrium fluctuations and extend our theory
to the case of contacts with arbitrary transparency.
Finally we investigate the suppression of fluctuations
in the small capacitance limit
due to charge screening.
\end{abstract}

\maketitle

The full counting statistics of current fluctuations in mesoscopic
conductors has attracted the attention of many theoretical
works during the last decade. In a pioneering paper
Levitov et al. explain the universality of current
statistics: Coherent charge transfer through a two-terminal
conductor can be seen as probabilistic
process governed by a set of transmission probabilities \cite{Levitov1}.
In the following several methods have been developed to
obtain the full counting statistics of mesoscopic conductors:
In the original work the authors used a fully quantum mechanical
approach and described the conductor by its unitary scattering
matrix \cite{Levitov2,Muzykantskii1,Levitov3}. 
Nazarov provided a description in terms of 
Keldysh-Green's functions that can be conveniently applied to
conductors with a large number of channels 
\cite{Nazarov1}. De Jong characterized
double barrier structures successfully by a fully
classical method \cite{DeJong1}
based on the exclusion principle. 
Nagaev proposed a semi-classical
diagrammatic scheme to obtain higher cumulants
in a systematic manner \cite{Nagaev1,Nagaev2}.
Recently, Pilgram et al. expressed the full counting statistics
in terms of a saddle point solution to a semi-classical stochastic
path integral \cite{Pilgram1}.

In parallel different works addressed the statistics
of quantities related to current such as the phases
in superconducting devices \cite{Belzig1,Belzig2}
or the momentum transfer
from electrons to a bent conductor
\cite{Kindermann1}. The interpretation
of the statistics of these quantities is less intuitive,
in general they cannot be characterized by transmission
probabilities only. Momentum transfer for example
is not quantized in contrast to charge transfer
\cite{Kindermann2}.


\begin{figure}[htb]
\begin{center}
\leavevmode
\psfig{file=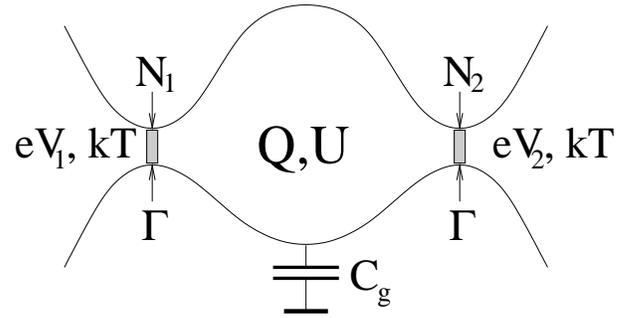,width=8cm}
\vspace{5mm}
\caption{Example of a mesoscopic conductor
considered in this article:
the chaotic cavity. We calculate the statistics
of the charge $Q$ inside the cavity. $U$ is the
electrostatic potential of the cavity, $V_1,V_2$
are the applied voltages, $N_1,N_2$ the number of
channels in the contacts, $\Gamma$ the contact transparency,
$kT$ the temperature, and $C_g$ the geometrical capacitance
of the cavity.
}
\label{Sketch System}
\end{center}
\end{figure}

In this article we report about a quantity
of fundamental interest in physics: the
fluctuations of charge inside a mesoscopic
conductor. As soon as parts of a mesoscopic
system are only coupled capacitively
(i.e. metallic gates close to a
mesoscopic conductor), charge fluctuations
play an important role: Decoherence for instance
is described as an average over a fluctuating
potential \cite{Decoherence1,Decoherence2} which then in
turn is coupled electrodynamically to
the charge. Charge correlations also
contribute to non-equilibrium decoherence
when a measurement process is taking place
\cite{Decoherence3,Decoherence4}. They are as well at
the origin of Coulomb drag effects \cite{Coulomb1}.
Such models often assume gaussian fluctuations,
which is justified in the weak coupling limit.
When coupling becomes stronger non-gaussian
fluctuations should be taken into account.
Methods to calculate higher cumulants are
therefore of great use!

In some sense charge fluctuations are 
diametrically opposed to the fluctuations
of current. For a current measurement we may
arbitrarily choose the cross section of the
conductor, for a charge measurement
we are forced to define exactly a volume
of charge we are interested in \cite{Volume1}.
The choice of this volume will necessarily
influence the statistics of charge
fluctuations. 
At this point it becomes clear that 
statistics of charge noise are in general
non-universal in contrast to current noise.
There is however a class of conductors
in which we expect universality to be preserved:
Disordered systems and in our case chaotic
cavities. In this case the charge of interest
can be characterized by one single parameter,
the dwell time of the electrons
inside the volume of the cavity \cite{Brouwer1}. 
Furthermore the hamiltonian of a chaotic
cavity is random, no basis nor any operator
is preferred.
Therefore the fluctuations
of any other internal operator such as the dipolar
moment will obey the same universal statistics as
the charge operator.

In this publication we present a method based
on random matrix theory \cite{Brouwer2} to
calculate the cumulants of charge noise in
chaotic cavities analytically order by order.
For the full generating function we show 
semi-analytical results that we confirmed numerically.
We emphasize that the example of the chaotic
cavity in the many channel limit
can be treated as well in the semi-classical
framework developed in Ref. \cite{Pilgram1}.
As we checked, the semi-classical results completely
agree with the calculation presented here.

In section \ref{Generating Function} we
derive the generating function of charge
fluctuations in terms of the scattering matrix. 
A similar expression for the second cumulant
has been derived long before
\cite{Buettiker2}. Results are in particular
available for the geometry considered in this
publication: the chaotic cavity \cite{Buettiker1}.
The charge-charge correlation
function has also been found at finite
frequencies \cite{Buettiker3}.
These works all emphasize the importance of
charge screening, interaction effects
have to be taken into account!
The charge in the cavity responds
to the fluctuations of the internal
electrostatic potential that is
conjugated to the charge.

We explain in detail the evaluation 
of the characteristic function for
an open cavity in equilibrium in section
\ref{Greensfunction Average}. We
present results for the non-equilibrium
case and contacts with arbitrary transparency
in section \ref{Results}.
We briefly discuss how to include interaction
effects into our calculation scheme.
As an application, we finally 
estimate the contribution of
non-gaussian fluctuations to dephasing
of electrons passing the chaotic cavity.

\section{Generating Function}
\label{Generating Function}

We consider a mesoscopic conductor as shown in
Fig.\ \ref{Sketch System}. The conductor is
described by a hamiltonian ${\bf \hat{H}_0}$.
The electrostatic potential $U$ on the conductor
can be varied. The full hamiltonian is therefore
${\bf \hat{H}} = {\bf \hat{H}_0} + U{\bf \hat{Q}}$
where ${\bf \hat{Q}}$ is the charge operator.
In analogy to current statistics \cite{Levitov3}
we define the following characteristic function
(we set $\hbar = 1$)
\begin{equation}
\label{Definition Characteristic Function}
\begin{array}{ll}
\chi_t(U) &= \langle e^{it({\bf \hat{H}_0}+\frac{U}{2}{\bf \hat{Q}})}
e^{-it({\bf \hat{H}_0}-\frac{U}{2}{\bf \hat{Q}})} \rangle\\
& = \langle {\bf \hat{T}_K}
e^{i\frac{U}{2}\int\!\!\!\int_0^t dt'(\pm){\bf \hat{Q}}(t')}
  \rangle.
\end{array}
\end{equation}
The first line of Eq. (\ref{Definition Characteristic Function})
in the Heisenberg picture will serve
as starting point for our calculation. The second line
in the interaction picture contains the Keldysh
time ordering operator ${\bf \hat{T}_K}$ and 
an integral $\int\!\!\!\int$ along the Keldysh contour 
from zero to $t$ and back again. 
The $\pm$ sign is positive on the upper Keldysh contour
and negative on the lower contour.
The interaction representation allows to
verify that the derivative 
$C_n = (-i)^n\partial^n \ln \chi(U)/ \partial U^n|_{U=0}$
generates the $n$th cumulant of charge fluctuations.

The Fourier transform of this characteristic function
is the probability distribution of charge integrated
over time $\int_0^t dt' Q(t')$, a quantity
without direct physical meaning.
Divided by the elementary charge $e$
it can be understood as the time spent by all electrons
in the cavity after time $t$. 
Alternatively it can be divided by a geometrical
capacitance $C_g$ to obtain the
probability distribution of the phase which is
the potential integrated over time $t$.
In this publication
we mainly concentrate on the characteristic function 
itself that generates the moments of zero frequency charge
fluctuations.

Using the procedure of Refs. \cite{Levitov2,Levitov3} the time evolution
operators defined in Eq.\ (\ref{Definition Characteristic Function})
can be expressed by the 
unitary scattering matrix of the mesoscopic
conductor and we obtain
\begin{equation}
\label{Intermediate Characteristic Function}
\begin{array}{c}
\ln \chi_t = \\ 
\quad\\
\frac{t}{2\pi} \int dE 
\mbox{Tr} 
 \ln \left({\bf 1 - n_E + n_E S^{\dagger}}(E,-\frac{U}{2})
{\bf S}(E,+\frac{U}{2}) \right).
\end{array}
\end{equation}
The energy integral reaches from the bottom of the band
to infinity. The matrix ${\bf S}(E,U)$ is the energy
and potential dependent scattering matrix that links incoming
and outgoing current amplitudes. The matrix ${\bf n_E}$
is diagonal and contains the Fermi occupation factors
$({\bf n_E})_{ii} = n_i(E) = [1+\exp((E-eV_i)/kT)]^{-1}$ 
of channel $i$ on its diagonal.
In practice it is useful to subtract the background
charge. We consider separately the
characteristic function at zero temperature and voltage
that can be expressed by the eigenvalues of the
scattering matrix $e^{i\phi_n}$
\begin{equation}
\label{Zero Characteristic Function}
\ln \chi_0(U) = 
i\frac{t}{2\pi} \int^{0} dE
\left\{ \sum_n\left(\phi_n(U/2) - \phi_n(-U/2)\right) \right\}.
\end{equation}
It is obvious from Eq.\ (\ref{Zero Characteristic Function})
that all even cumulants vanish at zero temperature and
voltage. The odd cumulants except the first vanish
after disorder averaging\cite{Disorder1}, because
$\langle \partial^k\phi_n / \partial U^k \rangle = 0$
for $k>1$. In the presence of disorder, 
subtracting $\ln \chi_0(U)$ from Eq.\
(\ref{Intermediate Characteristic Function})
is thus equivalent to subtracting the background charge.
We arrive at the following symmetric expression for
the characteristic function
\begin{equation}
\label{Characteristic Function}
\begin{array}{c}
\ln \chi  = \\
\quad\\
  \frac{t}{2\pi} \int_{-\infty}^{0} dE
\mbox{Tr}  \ln
\left[{\bf n_E + (1-n_E) S^{\dagger}}(E,\frac{U}{2})
{\bf S}(E,-\frac{U}{2}) \right]
\\
\quad\\
+ \frac{t}{2\pi} \int_{0}^{\infty} dE
\mbox{Tr}  \ln
\left[{\bf (1-n_E) + n_E S^{\dagger}}(E,-\frac{U}{2})
{\bf S}(E,\frac{U}{2}) \right]
.
\end{array}
\end{equation}
At equilibrium one can easily check that Eq.\
(\ref{Characteristic Function}) generates only
even cumulants. The characteristic function
then only depends on the eigenvalues $e^{i\phi_n}$
of the scattering matrix and is symmetric
with respect to $U \mapsto -U$.

\section{Average for a Chaotic Cavity}
\label{Greensfunction Average}

So far we have been very general. Eq.\ 
(\ref{Characteristic Function}) can be
applied to any mesoscopic conductor, as soon as
its scattering matrix is known \cite{Comment1}.
We now investigate more closely
the generic example of
a many mode chaotic cavity
(we neglect weak localization corrections).
In this case the average of products
of at most four
energy (potential) dependent scattering matrices 
is known \cite{Brouwer1,Weidenmueller1}.
It remains nevertheless a non-trivial problem
to calculate the average of the
logarithms in Eq.\ (\ref{Characteristic Function})
that contain infinitely high powers of 
scattering matrices. We solve this
problem in two steps. First we
express the characteristic function
through a Green's function. We then
calculate the average of this Green's function
using a method developed in Ref. 
\cite{Brouwer2}. 

We define the following Green's functions
\begin{equation}
\label{Definition Greensfunction}
{\bf F}^{eq}(z,U) = \langle
\frac{1}{z- {\bf A}^{eq}(U)} \rangle,
{\bf F}^{neq}(z,U) = \langle
\frac{1}{z- {\bf A}^{neq}(U)} \rangle.
\end{equation}
The brackets denote the disorder average.
The matrices ${\bf A}$ abbreviate products of
scattering matrices
\begin{equation}
\begin{array}{c}
{\bf A}^{eq} {\bf = S^{\dagger}}(-U/2)
{\bf S}(U/2)\\
\quad\\
{\bf A}^{neq} {\bf = S_{11}^{\dagger}}
(-U/2){\bf S_{11}}(U/2) 
+ {\bf S_{12}^{\dagger}}(-U/2){\bf S_{12}}(U/2).
\end{array}
\end{equation}
The dimension of the full scattering matrix
is $N\times N$, its block ${\bf S_{11}}$
(the reflection matrix)
measures $N_1 \times N_1$.
We introduce the temperature $kT$, the
applied voltage $V=V_1-V_2>0$ and the integration time $t$.
The characteristic functions at
equilibrium $(kT\gg eV)$ and for transport 
$(kT\ll eV)$ can then
be written as integrals over the variable $z$
of the Green's functions
\begin{equation}
\label{Characteristic Through Greensfunction}
\begin{array}{c}
\chi^{eq}(U) =
 \frac{tkT}{2\pi}
\int_1^{\infty}dz \frac{\ln(z)}{1+z}
\left(2N + (1+z) \right.\\
\quad\\
\left.
\mbox{Tr}
\left({\bf F}^{eq}(-z,U) + {\bf F}^{eq}(-z,-U)
\right)\right),
\end{array}
\end{equation}
and
\begin{equation}
\begin{array}{c}
\chi^{neq}(U) = \frac{teV}{2\pi}
\int_0^{\infty} \frac{dz}{1+z}
\\
\quad\\
\left(N_1 + (1+z)
\mbox{Tr}
{\bf F}^{neq}(-z,U)\right).
\end{array}
\end{equation}
To outline the principles of the calculations
we first specially discuss the simplest
case of an open cavity at equilibrium.
As Brouwer and B\"uttiker
\cite{Brouwer1} we introduce the
potential dependence of the scattering matrix
via a virtual stub described by a 
$M\times M$ reflection matrix
\begin{equation}
\label{Stub Definition}
{\bf r_s}(U) = -{\bf e}^{ieU{\bf \Phi}} \qquad \phi = \mbox{Tr} {\bf \Phi}.
\end{equation}
The trace $\phi$ is linked to the mean level density by
$\phi = 2\pi dn/(edU) = 2\pi N_F$ and to the dwell time by
$\phi = G\tau$ where $G$ is the total dimensionless conductance into
the cavity and $e$ the elementary charge. 
The stub is assumed to be very large compared
to the exits of the cavity; then
the precise structure of the matrix ${\bf \Phi}$
becomes unimportant for the calculation.
The total scattering matrix of the chaotic cavity may be
expressed as geometrical sum
\begin{equation}
\label{Scattering Matrix Sum}
{\bf S}(U) = {\bf U_{aa} + U_{ab}}
\left({\bf 1-r_s}(U)
{\bf U_{bb}}
\right)^{-1} {\bf r_s}(U) {\bf U_{ba}}.
\end{equation}
For an open chaotic cavity the 
$(N+M)\times (N+M)$ matrix
${\bf U}$ is distributed according to the
circular ensemble of unitary matrices. 
It is divided in blocks ${\bf U_{aa}}$
of size $N\times N$ and ${\bf U_{bb}}$
of size $M\times M$.
The average over this ensemble can be carried out by means
of a diagrammatic technique \cite{Brouwer2}.
The technical details
of this average are presented in appendix 
\ref{Averaging}.
The trace of the Green's function turns out to be
\begin{equation}
\label{Greensfunction Result}
\mbox{Tr} {\bf F}^{eq} = N \frac{1}{z-y(z,U)}
\end{equation}
where $y$ is the root of a cubic Dyson equation
\begin{equation}
\label{Dyson Equation Result}
z - z\left(1-ieU\tau\right)y -\left(1+ieU\tau\right)y^2 +  y^3
= 0.
\end{equation}

\begin{figure}[t]
\begin{center}
\leavevmode
\includegraphics[width=80mm]{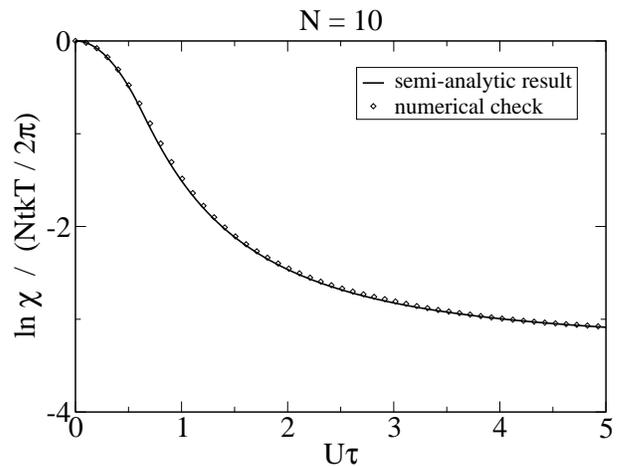}
\vspace{3mm}

\caption{Comparison of semi-analytical and numerical result for the
  logarithm of the characteristic function of equilibrium charge
  fluctuations in an open cavity. We choose $M=160$
  for the size of the numerical random hamiltonian
  describing the cavity.
}
\label{Characteristic Function Plot}
\end{center}
\end{figure}

Unfortunately the analytic solution of this
polynomial equation is rather cumbersome.
But we have two ways to continue. If we are
only interested in the first few cumulants
of charge fluctuations, we may expand the
root $y$ up to a certain order in the potential $U$
\begin{equation}
y = y_0 + y_1U + y_2U^2 + y_3U^3 + \dots
\end{equation}
and then solve the cubic equation order by
order
(As a side product of our calculation
we can obtain in the same way the average 
$\mbox{Tr}\langle({\bf S^{\dagger}}(-U/2){\bf S}(U/2))^n\rangle$ by expanding
$y$ in powers of $1/z$).
We then obtain the following expansion
of the characteristic function
\begin{equation}
\begin{array}{c}
\ln\chi^{eq}(U) =\\
\quad\\
 N_1\frac{tkT}{2\pi}
\left\{
-(eU\tau)^2 + \frac{1}{12}(eU\tau)^4 - \frac{1}{45}(eU\tau)^6 + \dots
\right\}.
\end{array}
\end{equation}
The equilibrium fluctuations of the charge in the open cavity
are non-gaussian. In the next section we will study
how this result depends on the back-reflection probability
$1-\Gamma$ of the leads.

If we are interested in the full characteristic function
we can find the root of Eq.\ (\ref{Dyson Equation Result})
numerically and then carry out the integration
in Eq.\ (\ref{Characteristic Through Greensfunction}).
The result is plotted in Fig.\ \ref{Characteristic Function Plot}.
The diamonds indicate a numerical check that was
performed by numerically averaging over a large set of
scattering matrices. This fully numerical solution
is very time consuming compared to finding roots of
the Dyson equation. The large $U$ limit of the
characteristic function can be determined analytically
to be $\ln \chi(U\rightarrow\infty) = -\pi NtkT/12$.
The characteristic function $\chi$ is therefore properly
behaved for large times $t \gg (NkT)^{-1}$. 
As discussed in Ref. \cite{Levitov3}
this is the range of validity of Eq.\ 
(\ref{Intermediate Characteristic Function}). 
Out of equilibrium the 
applied voltage plays the role of the temperature
$t \gg (NeV)^{-1}$.

\section{Results}
\label{Results}

The generalization of our approach to cavities with
back-reflection at the contacts is
not completely straightforward. On the one hand
the scattering matrix as given in Eq.\ 
(\ref{Scattering Matrix Sum}) must be extended
and its average $\langle {\bf S}(U) \rangle$
is no longer zero. This complication modifies
the averaging procedure for the Green's functions.
We discuss the crucial points in appendix 
\ref{Cavity With Barrier}. On the other hand
the calculation becomes lengthy and requires
some algebraic software.

We use the parameters $N = N_1 + N_2$ for
the total number of channels and
$\lambda = (N_1-N_2)/N$ to describe the
asymmetry of the cavity (see Fig.\ \ref{Sketch System}). 
The transmission probability to pass one of the
contacts of the cavity is denoted
by $\Gamma$. For simplicity we choose
it to be the same in both contacts.

For the cumulants of fluctuations
at equilibrium we obtain
\begin{equation}
\label{Equilibrium Results}
\begin{array}{ll}
C_2^{eq} & = N\Gamma\frac{kTt}{\pi} (e\tau)^2 \\
\quad\\
C_4^{eq} & = N\Gamma\frac{kTt}{\pi} (e\tau)^4 (2 - \Gamma)\\
\quad\\
C_6^{eq} & = N\Gamma\frac{kTt}{\pi} (e\tau)^6 
(24 - 24\Gamma + 8\Gamma^2).\\
\end{array}
\end{equation}
The dwell time is given by $\tau$ and $e$ is the unity
charge. 
This result is universal in the sense that it describes
also the fluctuations of any other operator inside the
cavity.
One has simply
to replace the constant $(e\tau)$ by other units.
It is interesting to note that
the higher cumulants increase with decreasing
transparency of the leads. We observe the same
behavior for non-equilibrium charge fluctuations and
for current fluctuations in chaotic cavities
\cite{Nagaev2}. Towards the limit of tunneling
contacts the non-gaussian fluctuations get stronger.

\begin{figure}[t]
\begin{center}
\leavevmode
\includegraphics[width=70mm]{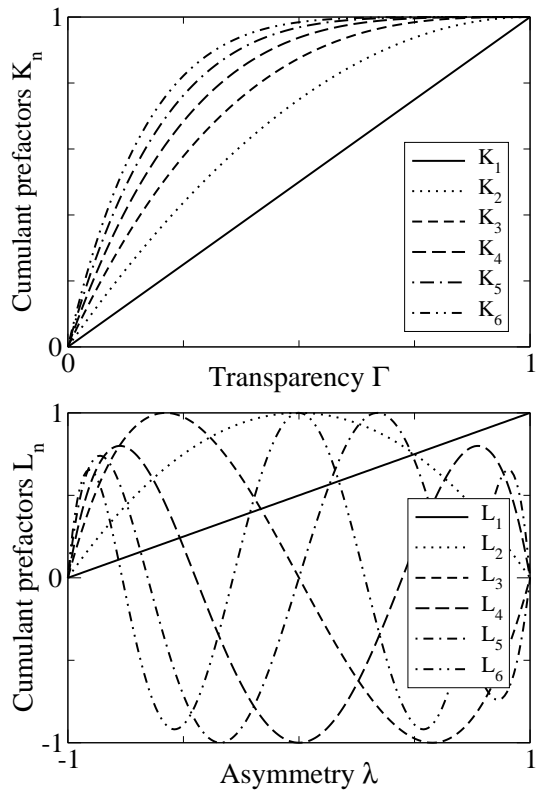}
\vspace{3mm}

\caption{
Parameter dependence of non-equilibrium cumulants
of charge noise.
{\bf Upper panel}: 
Each cumulant $C_n$ depends on transmission $\Gamma$. 
The prefactors of low cumulants tend to zero 
in the tunneling limit
whereas high cumulants stay important.
{\bf Lower panel}:
Each cumulant depends on the asymmetry 
$\lambda = (N_1-N_2)/(N_1+N_2)$. Cumulants higher
than two may change their sign.
}
\label{Transparency Plot}
\end{center}
\end{figure}

In the transport regime 
the characteristic function
can be written in the following way
\begin{equation}
\label{Nonequilibrium Series}
\ln \chi^{neq} =  \frac{NteV}{2\pi}
\sum_n K_n(\Gamma) L_n(\lambda) (ieU\tau)^n.
\end{equation}
The functions $K_n$ and $L_n$ describe the dependence
of the cumulants on transparency of the contacts and
asymmetry of the cavity. The first few coefficients
are given by
\begin{equation}
\label{Transparency Factors}
\begin{array}{l}
K_1  = \Gamma
\qquad
K_2  = \Gamma(2-\Gamma)
\qquad
K_3  = \Gamma(3-3\Gamma+\Gamma^2)\\
\quad\\
\quad\\
L_1  = \frac{1+\lambda}{2}
\qquad
L_2  = \frac{1-\lambda^2}{8}
\qquad
L_3  = -\frac{\lambda (1-\lambda^2)}{12}
\end{array}
\end{equation}
These factors are plotted in Fig.\ \ref{Transparency Plot} as a function
of transparency and asymmetry. The higher
the cumulant, the slower the factors $K_n$ diminish towards the tunneling
limit. The plot indicates a non-analytic behavior for $\Gamma\rightarrow
0$ and $n\rightarrow \infty$. A similar effect occurs in
extremely asymmetric cavities at $|\lambda|\simeq 1$: Non-gaussian
fluctuations are enhanced.

Up to now we considered a completely non-interacting system 
($C_g \gg e^2N_F$). We now relax the condition
$C_g \gg e^2N_F$ by including the effect of charge screening.
In the many-channel limit considered here
Coulomb blockade effects are unimportant.
Interaction effects can then be treated by
including a self-consistent response of the
electrostatic potential $U$ (the bottom of the band)
to the charge fluctuations
\cite{Buettiker1}.
We introduce a fluctuating 
electrostatic potential $U=U_0 + \delta U(t)$
that is coupled capacitively to the fluctuations
of the total screened charge on the cavity 
$\delta Q_{scr} = C_g\delta U$.
We then have to include the response of
charge cumulants $C_n$ to fluctuations of the
potential $\delta U$, the so called cascade corrections
\cite{Nagaev1,Nagaev2}. In the case of 
cavities this procedure is almost trivial, because only
the first cumulant of charge $C_1$, the mean charge,
depends on the electrostatic potential, see
appendix \ref{Screened Charge Fluctuations}.
It turns out that the screened
cumulants differ from the unscreened cumulants
by a universal prefactor (in the
linear bias regime)
\begin{equation}
\label{Screened Fluctuations}
C_n^{scr} = \left(1+\frac{e^2N_F}{C_g}\right)^{-n}C_n.
\end{equation}
An identical prefactor has been obtained in Ref.
\cite{Buettiker1} for the second cumulant. 
This simple result relies on
two facts: All cumulants higher than one do not
depend on the electrostatic potential, because
correlators of scattering matrices
$\langle (S^{\dagger}(U_1)S(U_2))^n\rangle$
only depend on the difference $U_1-U_2$ after disorder
averaging.
Furthermore, the density of states $N_F$ depends
in principle on the par\-ti\-cu\-lar scattering matrix.
Only in the many channel limit we can consider
$N_F$ and $C_n$ as being independent when averaging
over disorder.
In general a relation as
simple as Eq.\ (\ref{Screened Fluctuations})
cannot be expected!

\section{Application: Phasefluctuations}

In this chapter we estimate the contribution
of non-gaussian fluctuations to the dephasing
of electrons passing a chaotic cavity.
This allows us to formulate a va\-li\-di\-ty condition
for the gaussian approximation.
Models for dephasing 
in chaotic cavities have been considered
for instance
in Refs. \cite{Baranger1,Brouwer3}.
Dephasing rates in cavities have been measured
by Huibers et al. \cite{Huibers1} and
Hackens et al. \cite{Hackens1}.
Dephasing due to charge fluctuations
has been treated in the scattering formalism
in Refs. \cite{Decoherence2,Seelig1,Buettiker4}. 

We approximate the dynamical phase picked up by an electron in 
the cavity by
\begin{equation}
\label{Dynamical Phase}
\phi \simeq e\int_0^{\tau} dt' U(t') = \frac{e}{C_g}\int_0^{\tau}
dt' Q^{scr}(t').
\end{equation}
The screened charge is denoted by $Q^{scr}$, its statistics
is in the long time limit
given by Eq.\ (\ref{Screened Fluctuations}).
We consider the phase $\phi$ to be a classically fluctuating
field \cite{Decoherence2} (with non-gaussian fluctuations)
and average the phase over equilibrium charge fluctuations
\begin{equation}
\label{Phase Average}
\langle e^{i\phi} \rangle  \simeq
\exp\left\{-\frac{1}{2}\frac{e^2}{C_g^2}C^{scr}_2 
+ \frac{1}{24}\frac{e^4}{C_g^4}C^{scr}_4 -\dots\right\}.
\end{equation}
Here we assumed additionally that fluctuations on
the time scale of $\tau$ behave
as fluctuations in the long time limit.
This is justified at high temperatures $kT \gg \tau^{-1}$
if the frequency dependence of the 
response function $\partial \langle Q \rangle / \partial U$
(see appendix \ref{Screened Charge Fluctuations}) may
be neglected.
We combine Eqs.\  (\ref{Equilibrium Results})
and (\ref{Screened Fluctuations}), and insert the
cumulants into Eq.\ (\ref{Phase Average}).
A comparison of second and forth cumulant
gives the following condition for the validity of the
gaussian approximation
\begin{equation}
\label{Dephasing Condition}
\frac{1}{G^2}\frac{C_{\mu}^2}{C_g^2}
\frac{2-\Gamma}{12} \ll 1
\end{equation}
where we introduced the electrochemical capacitance
$C_{\mu}^{-1} = C_g^{-1} + (e^2N_F)^{-1}$ and the
dimensionless conductance $G=\Gamma N$.
Note that $C_{\mu} < C_g$, therefore condition
(\ref{Dephasing Condition}) always holds
in the large conductance limit considered in this article.
Nevertheless, Eq.\ (\ref{Dephasing Condition})
indicates that non-gaussian corrections 
become important in the limit of few channels.
They are strongest in the
charge neutral limit $C_{\mu} = C_g$ and increase
in the tunneling limit $\Gamma \rightarrow 0$.

\section{Conclusions}

In this article we described a way to calculate higher
cumulants of charge fluctuations inside a mesoscopic
conductor in the long-time limit. 
We first derived a general generating functional
for such fluctuations in terms of the scattering matrix
of a non-interacting mesoscopic system.
As a specific example we considered a chaotic cavity and showed
how to average the generating functional over disorder using
random matrix theory.
We then studied the dependence of the generating functional
on parameters as the asymmetry or transparency of the leads
attached to the cavity.
We find that higher cumulants contribute most to fluctuations
in the tunneling limit.
Finally we calculated the suppression of  fluctuations
in the presence of  screening. 
Although higher cumulants of charge fluctuations are
probably not directly measurable, their knowledge is
of great practical importance. It allows for instance
to study decoherence effects due to fluctuations beyond the
gaussian theory.

The authors thank P. Samuelsson and E. V. Sukhorukov
for valuable discussions and acknowledge the support
of the Swiss National Science Foundation.

\appendix

\section{Disorder Average of Green's functions}
\label{Averaging}

In this appendix we discuss how to average
the potential dependent Green's functions 
(\ref{Definition Greensfunction}) in the
equilibrium case. 
In Ref. \cite{Brouwer2} Brouwer and Beenakker present a procedure to
calculate the distribution of transmission eigenvalues
of chaotic cavities from a disorder averaged
Green's function.
We adapt this procedure 
to our purposes. In a first step the series
in Eq.\ (\ref{Scattering Matrix Sum})
is reexpressed in a compact formulation
by introducing a set of matrices
\begin{equation}
{
\begin{array}{c}
\begin{array}{ll}
{\bf \tilde{S}} = 
\left(\begin{array}{cc} {\bf S}(U/2) & 0 \\ 0 & {\bf S^{\dagger}}(-U/2)\end{array}\right), &
{\bf C} = 
\left(\begin{array}{cc} 0 & {\bf 1} \\ {\bf 1} & 0\end{array}\right), 
\end{array}\\
\quad\\
\begin{array}{ll}
{\bf \tilde{F}}(z) = 
\left(\begin{array}{cc} 0 & {\bf F'}(z) \\ {\bf F}(z) &
  0\end{array}\right), &
{\bf T'} = \left(\begin{array}{cccc} {\bf t} & 0 & 0 & 0\\
                 0 & 0 & {\bf t^{\dagger}} & 0\end{array}\right),
\end{array}\\
\quad\\
\begin{array}{ll}
{\bf T} =  {\bf T'}^T, &
{\bf R'} = 
\left(\begin{array}{cccc} {\bf r'} & 0 & 0 & 0 \\ 0 & {\bf r_s}(U/2) & 0 & 0 \\
      0 & 0 & {\bf (r')^{\dagger}} & 0 \\ 
      0 & 0 & 0 & {\bf r_s^{\dagger}}(-U/2)\end{array}\right).
\end{array}
\end{array}
}
\end{equation}
The transmission and reflection 
matrices ${\bf t},{\bf r}$ and ${\bf r'}$ describe the back-reflection
in the contacts. Without barriers we have ${\bf t} = {\bf 1}$
and ${\bf r} = {\bf r'} = {\bf 0}$.
The fluctuating part of the scattering matrix then is
\begin{equation}
{
{\bf \delta\tilde{S}} = {\bf T'}\left(1-{\bf \tilde{U}R'}\right)^{-1} {\bf
  \tilde{U}T},
\quad {\bf \tilde{U}} = 
\left(\begin{array}{cc} {\bf U} & 0 \\ 0 & {\bf U^{\dagger}}\end{array}\right)
}
\end{equation}
where ${\bf U}$ is the uniformly distributed unitary random matrix.
The matrix Green's function ${\bf \tilde{F}}$
that contains the desired Green's function 
(\ref{Definition Greensfunction}) as
off-diagonal element can be written as difference
${\bf \tilde{F}} = {\bf \tilde{F}^+}-{\bf \tilde{F}^-}$.
Both components
\begin{equation}
\label{Matrix Greensfunction}
{\bf \tilde{F}^{\pm}} = \frac{1}{2\sqrt{z}} {\bf C}\frac{1}{\sqrt{z} 
\mp {\bf {\tilde{S}C}}}
\end{equation}
can be averaged separately. According to Brouwer and Beenakker
\cite{Brouwer2} they obey a Dyson equation
\begin{equation}
{\bf \tilde{F}^{\pm}} = {\bf X^{\pm}} 
\left( 1 + {\bf \Sigma^{\pm}(\tilde{F}^{\pm})\tilde{F}^{\pm}}\right)
\end{equation}
with a self-energy matrix ${\bf \Sigma^{\pm}} = {\bf \Sigma^{\pm}}
({\bf \tilde{F}^{\pm}})$
depending on the Green's function ${\bf \tilde{F}^{\pm}}$. 
The matrix ${\bf X^{\pm}}$
is defined as ${\bf X} = {\bf R' + TCT'}z^{-1/2}$.
The equation for the self-energy may now be expanded
up to first order in $1/M$ where $M$ is the dimension of the
virtual potential dependent stub (\ref{Stub Definition}).
After impurity averaging and
some algebra we obtain that the self-energy is of
the form
\begin{equation}
{\bf \Sigma^{\pm}} = \pm \frac{y}{\sqrt{z}}
\left(
\begin{array}{cc}
&1\\
1&
\end{array}
\right)
\end{equation}
and fulfills Eq.\ (\ref{Dyson Equation Result}).

\section{Extension to Cavities with Barriers} 
\label{Cavity With Barrier}

In the presence of back-reflection at the contacts
we have to modify
the summation carried out by Brouwer and Beenakker
\cite{Brouwer2}.
Now the scattering matrix of the leads is 
given by ${\bf t} = i\sqrt{\Gamma}$ and
${\bf r} = {\bf r'} = \sqrt{1-\Gamma}$.
The additional complication due to barriers
arises from the fact that the mean of the scattering
matrix ${\bf \bar{S}}$ is no longer zero. 
The needed form of the matrix ${\bf C}$ is
\begin{equation}
{\bf C} = \left(\begin{array}{cc}
0 & {\bf C_1}\\
{\bf C_2} & 0\\
\end{array}\right).
\end{equation}
In order to obtain the density of transmission
eigenvalues as in Ref. \cite{Brouwer2} one chooses
\begin{equation}
\begin{array}{c}
\bf{C_1} = 
\left( \begin{array}{cc} {\bf 1} & 0\\0 & 0\end{array}\right)
\qquad
\bf{C_2} = 
\left( \begin{array}{cc}0 & 0\\0 & {\bf 1}\end{array}\right)\\
\quad
\end{array}
\end{equation}
and exploits the fact that 
${\bf C_1},{\bf C_2},{\bf \bar{S}}$ commute and
${\bf C_1C_2}=0$. 
These conditions do not hold in our case.
The matrix ${\bf C}$ contains more entries.
In equilibrium for
instance we need  ${\bf C_1} = {\bf C_2} = 1$.
We overcome this problem by replacing ${\bf C}$ through a matrix 
${\bf C'}$ that encounters the back-reflection in the 
leads
\begin{equation}
{\bf C'} = {\bf C}\left( {\bf 1 -RC}z^{-1/2}\right)^{-1}
\quad {\bf R} = \left(\begin{array}{cc}
{\bf r} & 0\\ 0 & {\bf r^{\dagger}}\end{array}\right).
\end{equation}
One can easily verify that this substitution generalizes the
Green's function introduced in Eq.\ (\ref{Matrix Greensfunction})
to the case of barriers at the contacts.
The rest of the calculation follows the lines of
appendix \ref{Averaging}.

\section{Screened Charge Fluctuations}
\label{Screened Charge Fluctuations}

In this appendix we explain the derivation of
Eq.\ (\ref{Screened Fluctuations}) using the cascade principle
(for an introduction to cascade corrections
we refer the reader to Refs. \cite{Nagaev1,Nagaev2,Pilgram1}).
Charge and potential fluctuations of the
cavity are linked via the capacitance $C_g$:
\begin{equation}
\label{Potential Fluctuations}
\delta Q_{scr} = C_g \delta U, \qquad
\delta Q_{scr} = \delta Q +
\frac{\partial \langle Q \rangle}{\partial U} 
\delta U.
\end{equation}
The charge fluctuations are composed of bare fluctuations
at constant potential $\delta Q$ 
and the linear response 
$\partial \langle Q \rangle/\partial U = -e^2N_F$
to $\delta U$ (the screening charge).
The fluctuations of $\delta Q$ 
are known from the non-interacting problem
(see Eqs. (\ref{Equilibrium Results}) and
(\ref{Nonequilibrium Series})).
Solving Eq. (\ref{Potential Fluctuations})
for the total charge fluctuations $\delta Q_{scr}$
the third cumulant without cascade corrections
may be written as
\begin{equation}
\langle (\delta Q_{scr})^3 \rangle =  
\left(1+\frac{e^2N_F}{C_g}\right)^{-3}
\langle (\delta Q)^3 \rangle.
\end{equation}
The two possible cascade correction are given by
\cite{Nagaev1}
\begin{equation}
3 \frac{\partial \langle (\delta Q_{scr})^2 \rangle}
{\partial U}
\langle \delta U \delta Q_{scr} \rangle, \qquad
3 \frac{\partial^2 \langle \delta Q_{scr} \rangle}
{\partial U^2}
\langle \delta U \delta Q_{scr} \rangle^2.
\end{equation}
But these corrections are zero, because 
$\langle (\delta Q_{scr})^2 \rangle$
does not depend on $U$ and $\langle \delta Q_{scr} \rangle$
depends linearly on $U$. Cascade corrections
to all higher cumulants vanish in the same way.
Eq.\ (\ref{Screened Fluctuations}) is therefore
valid for arbitrary $n$.


\begin{thebibliography}{02}
\bibitem{Volume1}       In principle, this volume can be of
                        microscopic dimensions. In practice,
                        we are often interested in charge
                        fluctuations associated with a capacitance,
                        and the volume is defined by the geometry
                        of the conductor.
\bibitem{Disorder1}     The first derivative 
                        $ \partial \phi_n/\partial U$ is proportional
                        to the density of states $N_F$
                        which does not depend on the potential
                        in a disordered system
                        (up to small corrections of the order
                        $eU/E_F$ where $E_F$ is the Fermi energy.
\bibitem{Comment1}      To achieve a Gauge invariant theory which
                        depends only on potential differences a
                        self-consistent treatment of the bottom
                        of the band in the energy integral of Eq.\ 
                        (\ref{Intermediate Characteristic Function})
                        is necessary. We return to this point in
                        section\ \ref{Results} where we discuss
                        how to include interaction effects.
\end{thebibliography}
\end{document}